\documentclass[reprint,aps,pre,floatfix]{revtex4-1}
\usepackage[T1]{fontenc}
\usepackage[utf8]{inputenc}
\usepackage{amsmath,amssymb,bm,graphicx,booktabs}
\usepackage[hidelinks]{hyperref}
\newcommand{\Ew}{\mathcal E}
\newcommand{\Pw}{\mathcal P}
\newcommand{\mean}[1]{\langle #1\rangle}
\newcommand{\Ree}{\operatorname{Re}}
\newcommand{\dd}{\mathrm d}
\begin{document}
\title{Amplitude-Dependent Damping of Gravity--Capillary Waves in Power-Law Fluids: An Energy-Dissipation Framework}
\author{Sabbir Hassan}
\affiliation{Department of Mechanical Engineering, Texas Tech University, Lubbock, Texas, USA}
\date{\today}

\begin{abstract}
Classical gravity--capillary theory locates the restoring-force crossover, but the wave amplitudes and depths for which energy losses remain measurable depend on the constitutive response and on where viscous dissipation occurs. We develop a weak-loss mechanical-energy model for small-steepness progressive surface waves in memoryless power--law liquids. The full invariant of the leading inviscid strain field yields the deep-water bulk law $\dd A/\dd t=-C_tA^n$, with $C_t=4Kk^2(2k\Omega)^{n-1}/[(n+1)\rho]$, and the corresponding spatial law $\dd A/\dd x=-(C_t/v_g)A^n$. For finite depth we calculate bulk dissipation using a phase- and depth-resolved quadrature $J_n(kh)$ and formulate a separate oscillatory power--law bottom layer. We solve the latter numerically in periodic state, obtaining dimensionless integrated wall losses $I_{0.6}\simeq0.5432$ and $I_{1.4}\simeq0.4708$ and recovering the exact Newtonian benchmark $I_1=1/2$. These values make it possible to evaluate conditional finite-depth bulk--wall dominance boundaries rather than only their amplitude exponents. The theory is organized into restoring-force, attenuation, local propagation-length, and dissipation-mechanism regime maps; at fixed wave steepness it additionally predicts a capillary-side maximum of normalized temporal damping for $0<n<2/3$. Grid, time-step, domain, and energy-work checks support the reduced wall calculation, while the coupled bulk--wall approximation still requires independent free-surface numerical or experimental validation. The results provide a reproducible single-fluid reference for subsequent two-liquid interfacial-wave modeling.

\end{abstract}
\maketitle

\section{Introduction}
\label{sec:intro}

\subsection{Physical context, practical motivation, and scope}

Surface gravity--capillary waves provide an accessible manifestation of the competition among inertia, restoring forces, and viscous dissipation. Gravity principally restores sufficiently long waves, whereas surface tension becomes increasingly important as wavelength decreases. The leading conservative dispersion relation therefore sets the oscillation frequency and propagation speeds, while viscous stresses determine how mechanical wave energy is converted into heat. This separation suggests a useful measurement principle: the observed wavelength and frequency constrain the conservative wave response, and the accompanying temporal or spatial attenuation carries additional information about the fluid and its boundaries. It is particularly attractive when immersion of a conventional rheometric probe is undesirable, the available liquid volume is small, or repeated, localized measurements are needed. Nevertheless, the separation is asymptotic rather than exact: substantial viscosity may alter dispersion and even suppress propagating modes, and a measured amplitude envelope can contain losses unrelated to the constitutive behavior of the liquid \cite{LeBlondMainardi1987,Behroozi2011}.

The relevance of short surface waves extends from environmental measurements to laboratory-scale fluid characterization. In air--sea interaction, gravity--capillary waves affect the short-wave spectrum used in remote sensing and are influenced by wind forcing and microbreaking, as illustrated by Laxague \emph{et al.} \cite{LaxagueEtal2018}. In small-volume measurements, optical methods can resolve capillary-wave motion without placing a mechanical sensor in contact with the specimen. Liu, Kijanka, and Urban demonstrated the use of optical coherence tomography (OCT) to measure capillary-wave phase velocities at air--liquid interfaces, including those of water, whole blood, and plasma \cite{LiuKijankaUrban2020}. Liu and Urban subsequently developed optical coherence viscometry, combining transient acoustic-radiation-force excitation with OCT measurements of wave dispersion and attenuation to infer the viscosity of Newtonian liquids, including water--glycerol mixtures and plasma \cite{LiuUrban2021OCV}. These studies establish both the practical promise and the measurement constraints of wave-based characterization: a finite observation window, limited displacement sensitivity, and the possibility of pronounced damping at short wavelengths. They do not, however, establish that a Newtonian attenuation inversion can be used unchanged for fluids with shear-dependent constitutive behavior.

That distinction matters for suspensions, polymer solutions, fluid-mud systems, and other soft or formulated liquids that can exhibit a viscosity varying with deformation rate. When an instantaneous generalized-Newtonian description is appropriate over the relevant shear-rate interval, wave-induced deformation changes not only with depth and wavelength but also with amplitude. Hence viscosity and dissipation are not fixed properties of a wave mode in the same sense as in the Newtonian limit. A larger-amplitude wave can sample a different range of strain rates from a smaller-amplitude wave of the same wavelength; its attenuation rate may therefore evolve as the wave decays. At finite depth, friction near a rigid bottom can contribute a distinct and potentially dominant loss, while localized excitation introduces geometrical spreading that must not be mistaken for constitutive attenuation. These considerations motivate a predictive connection between a specified rheological law, the full wave-induced deformation field, and the measurable amplitude envelope, with explicit limits on when a reduced wave model can be trusted.

\subsection{Background and related literature}

\paragraph*{Classical viscous-wave theory and energy transport.}
The Newtonian reference problem has a long and well-developed history. In \emph{Hydrodynamics}, Lamb derived the classical weak-viscosity decay of small-amplitude waves, including the leading deep-water amplitude rate $\Gamma_N=2\nu k^2$ \cite{Lamb1932}. This benchmark may be recovered either from the linearized viscous-wave problem or from the dissipation of the leading wave motion; it is not an exact attenuation law at arbitrary viscosity. Prosperetti addressed a different issue by solving the initial-value problem for small-amplitude standing waves on a deep viscous liquid. That study showed that transient motion can differ appreciably from either an irrotational short-time approximation or a discrete normal-mode description before the relevant asymptotic regime is reached \cite{Prosperetti1976}. LeBlond and Mainardi obtained complete linear solutions for viscous capillary--gravity oscillations at finite depth and examined the limits of wave propagation, including purely damped motions beyond the propagating branches \cite{LeBlondMainardi1987}. Together, these works establish both the Newtonian weak-loss baseline and the need to delimit its validity.

Subsequent studies refined how that baseline is used. Dias, Dyachenko, and Zakharov derived a weakly viscous potential-flow formulation from the linearized Navier--Stokes equations, introducing viscous corrections to both the dynamic and kinematic free-surface conditions; their treatment provides a systematic justification for a commonly used damped wave-envelope description \cite{DiasDyachenkoZakharov2008}. Antuono and Colagrossi solved the linearized finite-depth Newtonian problem with a no-slip bed, obtained approximate damping and celerity expressions beyond the deep-water, high-Reynolds-number limit, and compared them with numerical simulations \cite{AntuonoColagrossi2013}. For capillary--gravity waves, Behroozi, Smith, and Even investigated viscosity-induced changes to the dispersion relation and measured wavelength shifts in water--glycerol mixtures, reinforcing that viscous effects cannot always be confined to an imaginary correction to frequency or wavenumber \cite{Behroozi2011}. Armaroli \emph{et al.} reviewed viscous dispersion approximations and derived dissipative corrections to nonlinear wave-envelope dynamics; their results distinguish leading linear damping from typically smaller viscosity-dependent nonlinear corrections in the regimes examined \cite{ArmaroliEtal2018}. The energy-transport interpretation is equally established: for a weakly damped, narrowband mode, the conservative wave-energy flux is governed at leading order by the group velocity, as developed by Whitham and Lighthill \cite{Whitham1961Group,Whitham1974,Lighthill1978}. Thus the conversion between temporal and spatial attenuation is not, by itself, a new physical principle.

The measured attenuation can nevertheless be influenced by more than distributed bulk viscosity. In a review of gravity--capillary wave dissipation and nonlinear wave interactions, Berhanu emphasized the importance of dissipative mechanisms in laboratory and field wave dynamics and their consequences for interpreting nonlinear wave phenomena \cite{Berhanu2022}. Similarly, the short-wave spectra examined by Laxague \emph{et al.} demonstrate the importance of the gravity--capillary range for wave growth and microbreaking \cite{LaxagueEtal2018}. These studies motivate the use of distinct regime descriptors: a restoring-force ratio identifies the gravity--capillary crossover, but it does not independently determine a viscous attenuation rate, propagation length, or measurement-validity boundary.

\paragraph*{Wave measurements and non-Newtonian constitutive behavior.}
Work on non-Newtonian wave-based characterization predates the present study. Allouche \emph{et al.} applied electrocapillarity to weakly concentrated carboxymethylcellulose and xanthan-gum solutions, exploiting the propagation and viscous attenuation of capillary--gravity waves to access very-low-shear viscosity and complement conventional rheometry \cite{AlloucheEtal2015}. Their experiments establish that wave measurements can probe shear-thinning liquids; accordingly, neither the use of surface waves for non-Newtonian characterization nor the observation of rheology-dependent attenuation is claimed as new here. The OCT investigations of Liu \emph{et al.} and Liu and Urban provide a complementary route to noncontact excitation and sensing, but their demonstrated viscosity inversion is specifically Newtonian \cite{LiuKijankaUrban2020,LiuUrban2021OCV}. The remaining modeling question for the present wave class is how a specified nonlinear constitutive relation modifies the dissipated power throughout the oscillatory deformation field and, consequently, the local attenuation measured over a finite amplitude range.

More recent studies underscore both the breadth of non-Newtonian wave phenomena and the importance of distinguishing wave configurations. Calvo, Tamburrino, and Falc\'on experimentally compared transient, nonlinear long-wave propagation in Newtonian and shear-thinning liquids, reporting changes in propagation speed and nonlinear attenuation relative to water \cite{CalvoEtal2026}. Their solitary-wave regime differs from the locally monochromatic, small-steepness gravity--capillary waves considered here; in particular, a rheology-dependent correction to propagation speed need not be negligible in the former configuration. Hu, Chen, and Li incorporated independently measured shear-dependent fluid-mud rheology into a physics-based wave-damping model and demonstrated agreement with laboratory and field observations \cite{HuChenLi2026}. Notably, they found that a Newtonian surrogate can reproduce a similar net attenuation while distributing the dissipated energy differently within the mud. This result makes clear why matching only the observed decay coefficient does not establish that the correct constitutive dissipation mechanism has been captured. The present work consequently focuses on explicitly evaluating the local tensorial strain invariant before depth and cycle averaging, rather than prescribing a single apparent viscosity for the entire wave field.

\paragraph*{Finite-depth boundaries and layered-wave extensions.}
A rigid bottom introduces a second modeling problem even when the outer wave field is accurately approximated by inviscid theory. The outer solution satisfies impermeability but generally violates no slip, requiring an oscillatory shear layer to connect the wave-driven near-bed velocity to the stationary boundary. Pritchard, McArdle, and Wilson examined the oscillatory Stokes problem for power-law fluids and developed semi-analytical solutions supported by numerical calculations \cite{PritchardMcArdleWilson2011}. They showed that shear-thinning responses can decay algebraically away from the oscillating wall, whereas shear-thickening responses may exhibit finite penetration, in marked contrast with the Newtonian exponential Stokes layer. Their work supplies the appropriate nonlinear boundary-layer foundation, but it does not directly determine how such a layer competes with depth-integrated gravity--capillary wave dissipation. In particular, substituting an amplitude-dependent apparent viscosity into a Newtonian Stokes thickness does not generally recover the nonlinear wall response.

A related decomposition of dissipation is familiar from stratified and interfacial waves. Dore developed viscous-damping theory for stratified wave flows \cite{Dore1969}. Troy and Koseff combined theory and laboratory measurements for progressive two-layer interfacial waves and distinguished energy losses in the homogeneous fluid interiors from those in bottom, sidewall, and interfacial boundary layers; sidewall friction was especially important in their laboratory geometry \cite{TroyKoseff2006}. More recently, Suswanth and Rajan formulated spatial attenuation in multilayer stratified systems through a viscous dispersion relation and complex wavenumbers, demonstrating the sensitivity of attenuation to layer configuration and velocity gradients \cite{SuswanthRajan2025}. These studies establish that partitioning wave dissipation among physical regions is not a new idea. They also motivate a longer-term extension of the present approach to two-fluid interfacial modes, for which each layer can possess a distinct constitutive law. That two-fluid problem is not addressed in this paper.

\subsection{Research gap, present contributions, and paper organization}

The literature therefore provides several complementary ingredients: classical Newtonian attenuation and energy transport; noncontact capillary-wave measurements; experimental evidence of non-Newtonian wave damping; and theoretical descriptions of nonlinear oscillatory wall layers. What is not supplied by those ingredients alone is the particular \emph{wave-resolved, weak-dissipation closure} investigated here for a single power-law liquid: an explicit connection between the complete, locally varying rate-of-strain tensor of an approximately monochromatic gravity--capillary wave, the constitutive parameters $(K,n)$, finite-depth bulk dissipation, nonlinear bottom-layer loss, and the resulting temporal and spatial amplitude evolution. The objective is not to claim a new gravity--capillary crossover or a new Stokes-layer equation, but to assess whether these established components can be combined into quantitative and experimentally distinguishable attenuation predictions without introducing an arbitrary spatially uniform effective viscosity.

We first evaluate the generalized-Newtonian dissipation $\Phi=K\dot\gamma^{n+1}$ using the leading wave kinematics and a depth-integrated mechanical-energy balance. In deep water this yields an explicit amplitude-dependent attenuation coefficient and, under weak loss, a generally nonexponential temporal and spatial envelope that recovers the classical Newtonian limit at $n=1$. We next retain the full phase and depth dependence of the finite-depth strain invariant through a constitutive quadrature $J_n(kh)$, rather than averaging shear before applying the nonlinear stress law. For a no-slip bottom, we formulate the complementary power-law oscillatory boundary-layer problem and determine its dimensionless dissipation integral $I_n$ numerically for representative shear-thinning and shear-thickening indices, with the exact Newtonian solution serving as a benchmark. The resulting reduced composite model predicts distinct amplitude powers for bulk and wall attenuation and permits quantitative, model-based estimates of where the dominant dissipative contribution changes.

The analysis is then recast in terms of physically separate regime coordinates: gravity versus capillary restoring-force balance, attenuation relative to the wave period, local attenuation length relative to wavelength, and the ratio of wall to bulk dissipation. This separation allows us to ask whether attenuation extrema, usable propagation windows, and equal-dissipation contours coincide with the classical gravity--capillary crossover. It also exposes a potential ambiguity for wave-based rheometry: different constitutive laws or competing dissipation mechanisms can yield similar attenuation under a single measurement condition, so amplitude and wavelength sweeps may be needed for identification. The displayed regime boundaries are predictions of a specified reduced model and, where relevant, depend on stated rheological calibrations and validity thresholds; they are not asserted to be universal phase transitions.

Our analysis is restricted to small-steepness, weakly damped waves in an incompressible, memoryless power-law fluid with a clean free surface. The leading wave field is conservative, and the bulk--wall decomposition additionally requires separation of the outer wave and inner wall scales. Numerically resolving the reduced nonlinear Stokes problem establishes the wall-loss coefficient within that model, not independent validation of the coupled free-surface approximation. Testing the predicted attenuation and mechanism boundaries against full generalized-Newtonian flow calculations or measurements remains necessary before quantitative rheological inversion can be claimed. A regularized constitutive model would also be needed where the ideal power law fails at very low shear rates or where viscoelastic memory matters.

The remainder of the paper is organized as follows. Section~\ref{sec:framework} introduces the fluid model, assumptions, and mechanical-energy formulation. Section~\ref{sec:newt} summarizes the conservative dispersion relations and Newtonian attenuation benchmarks. Section~\ref{sec:deepPL} derives the deep-water power-law dissipation and amplitude laws, while Sec.~\ref{sec:finite} develops the finite-depth bulk quadrature and nonlinear bottom-layer calculation. Section~\ref{sec:regimemap} introduces the gravity--capillary--damping coordinates, and Sec.~\ref{sec:results} interprets the analytical and numerical predictions through the attenuation envelopes and regime maps. Finally, Sec.~\ref{sec:validity} discusses applicability, unresolved validation requirements, and the extension to two-fluid interfacial waves, followed by the conclusions in Sec.~\ref{sec:conclusions}.

\section{Formulation and mechanical-energy method}
\label{sec:framework}
\subsection{Physical problem and distinguished limits}
An incompressible liquid of density $\rho$ lies below a clean horizontal air--liquid interface $z=\eta(x,t)$; $z=0$ is the equilibrium surface and $x$ is the propagation direction. Deep water occupies $z<0$. In finite depth the bottom is stationary at $z=-h$. Surface tension $\sigma$ is constant, the air dynamic stress is neglected, and the flow is taken two-dimensional and approximately monochromatic,
\begin{equation}
\eta(x,t)\simeq A(x,t)\cos\theta,\quad
\theta=k_Rx-\omega_dt,\quad k_RA\ll1.
\label{eq:waveansatz}
\end{equation}
For a temporal free-decay mode we instead prescribe real $k$ and solve for complex $\omega$; for a driven spatial mode the driving frequency $\omega_d$ is real and the wavenumber is $k=k_R+i\alpha$. The symbol $k$ below denotes the leading real wavenumber unless complex continuation is explicitly indicated.

The liquid has Newtonian stress $\bm\tau=2\mu\mathbf D$ or generalized Newtonian power-law stress
\begin{equation}
\begin{aligned}
\bm\tau&=2K\dot\gamma^{\,n-1}\mathbf D,\qquad
\dot\gamma=(2\mathbf D:\mathbf D)^{1/2},\\
\mathbf D&=\tfrac12[\nabla\mathbf v+(\nabla\mathbf v)^T].
\end{aligned}
\label{eq:constitutive}
\end{equation}
Here $K$ has units Pa\,s$^n$, $n>0$ is the flow index, and $\mathbf v=(u,w)$. Setting $n=1$, $K=\mu$ recovers Newtonian stress. The pure power law ignores relaxation, yield stress, interfacial films, and finite low- and high-shear viscosity plateaus; its extrapolation to vanishing shear is particularly problematic for $n<1$.

We assume weak total amplitude damping $\Gamma/\Omega\ll1$ and weak spatial loss $\alpha/k_R\ll1$, together with small steepness. The leading wave kinematics and energy are therefore adopted from \emph{inviscid} linear theory, and nonlinear constitutive losses are treated perturbatively. This is not a direct solution of the nonlinear generalized-Newtonian free-boundary momentum equations. For a rigid bottom, additional requirements are $k\delta_n\ll1$, $\delta_n/h\ll1$, and a sufficiently localized wall correction (especially nontrivial for shear-thinning tails).

\subsection{From fluid mechanics to wave mechanical energy}
The relevant balance is the \emph{mechanical} portion of energy conservation, not the fluid's full thermodynamic total-energy equation. Dotting the incompressible momentum equation with $\mathbf v$, and adding gravitational potential energy, gives
\begin{equation}
\frac{\partial e_m}{\partial t}+\nabla\cdot\mathbf J_m=-\Phi,
\quad
\begin{aligned}
 e_m&=\tfrac12\rho|\mathbf v|^2+\rho gz,\\
 \mathbf J_m&=(e_m+p)\mathbf v-\bm\tau\cdot\mathbf v,\\
 \Phi&=\bm\tau:\mathbf D\ge0.
\end{aligned}
\label{eq:localenergy}
\end{equation}
The term $-\Phi$ is a mechanical-energy sink and a corresponding source of internal energy: total energy is not lost. Integrating through the water column with the moving-boundary contributions, including gravitational and capillary surface-energy storage, and averaging over a period leads to the wave balance
\begin{equation}
\boxed{\frac{\partial\Ew}{\partial t}
+\frac{\partial\mathcal F}{\partial x}=-\Pw},
\qquad
\Pw=\int_{\rm liquid}\mean{\Phi}\,\dd z.
\label{eq:wavebalance}
\end{equation}
Here $\Ew$ is mean mechanical energy per unit horizontal area (J\,m$^{-2}$), $\mathcal F$ is energy flux per unit transverse width (W\,m$^{-1}$), and $\Pw$ is power dissipated per unit area (W\,m$^{-2}$). The leading conservative flux is $\mathcal F\simeq v_g\Ew$ for locally narrowband waves \cite{Whitham1974,Lighthill1978}. A temporal uniform wavetrain satisfies $\dd\Ew/\dd t=-\Pw$; a steady, plane, right-moving wavetrain satisfies $v_g\dd\Ew/\dd x=-\Pw$. Once $\Ew=c_EA^2$, these give the common relations
\begin{equation}
\boxed{\frac{\dd A}{\dd t}=-\frac{\Pw(A)}{2c_EA},\qquad
\frac{\dd A}{\dd x}=-\frac{\Pw(A)}{2c_Ev_gA}}.
\label{eq:master}
\end{equation}
The second relation is a \emph{local plane-wave} transport law, not a replacement for cylindrical spreading in localized measurements. The regime coordinates introduced in Sec.~\ref{sec:regimemap} are consequently defined for local plane-wave attenuation, with radial geometric spreading considered separately. Unlike a complex-root calculation, Eq.~\eqref{eq:master} can also be evaluated when dissipation depends nonlinearly on $A$, provided the leading-wave approximation remains self-consistent.

\section{Conservative waves and Newtonian benchmark}
\label{sec:newt}
For a single clean surface, the known conservative dispersion relations are \cite{Lamb1932,Lighthill1978}
\begin{equation}
\Omega^2=gk+sk^3,\qquad
\Omega_h^2=(gk+sk^3)\tanh(kh),\qquad s=\frac\sigma\rho.
\label{eq:disp}
\end{equation}
We do not rederive the inviscid free-boundary problem. Differentiation gives the group velocities
\begin{align}
v_g&=\frac{g+3sk^2}{2\Omega},\label{eq:vg}\\
v_g^{(h)}&=\frac{(g+3sk^2)\tanh(kh)
+h(gk+sk^3)\operatorname{sech}^2(kh)}{2\Omega_h}.
\label{eq:vgh}
\end{align}
The corresponding crest speed is $c_p=\Omega/k$ and $c_p^{(h)}=\Omega_h/k$; the energy flux uses $v_g$, not $c_p$. The capillary length $\ell_c=\sqrt{s/g}$ locates the restoring-force crossover at $k_c=1/\ell_c$. For deep water $c_p^2=g/k+sk$ is minimized at $k_c$, where $c_{p,\min}=(4gs)^{1/4}$ and $v_g=c_p$. These are conservative reference quantities, not assertions that finite-viscosity wave speeds are exactly independent of rheology \cite{Behroozi2011,CalvoEtal2026}.

The deep-water Newtonian solution of the linearized Navier--Stokes eigenproblem has a potential and a rotational part; enforcing clean-surface kinematic and stress conditions yields the classical weak-viscosity eigenfrequency \cite{Lamb1932,Prosperetti1976,LeBlondMainardi1987}
\begin{equation}
\omega=\Omega-2i\nu k^2+o(\nu k^2),\qquad
\Gamma_N=2\nu k^2,\qquad \nu=\mu/\rho.
\label{eq:newtonian}
\end{equation}
The energy method independently returns the same leading rate. For the deep potential velocities
\begin{equation}
u=\Omega Ae^{kz}\cos\theta,\qquad
w=\Omega Ae^{kz}\sin\theta,
\label{eq:deepfield}
\end{equation}
the tensor invariant is $\mathbf D:\mathbf D=2k^2\Omega^2A^2e^{2kz}$. Consequently
\begin{equation}
\Pw_N=\int_{-\infty}^0 2\mu\mathbf D:\mathbf D\,\dd z
=2\mu k\Omega^2A^2,
\quad
\Ew=\frac{\rho\Omega^2A^2}{2k}.
\label{eq:newtonenergy}
\end{equation}
Using Eq.~\eqref{eq:master} yields $\Gamma_N=\Pw_N/(2\Ew)=2\nu k^2$ and $\alpha_N=2\nu k_R^2/v_g(k_R)$. Here $\Gamma_N$ is an amplitude-rate coefficient; the corresponding \emph{energy} decays twice as rapidly in logarithmic terms.

In finite depth, the leading distributed Newtonian bulk term remains $2\nu k^2$ within the outer-potential approximation, but a rigid \emph{no-slip} bottom adds an oscillatory wall loss. Its thin-layer asymptote is \cite{Lamb1932,AntuonoColagrossi2013}
\begin{equation}
\boxed{\Gamma_{N,h}\simeq 2\nu k^2+
\frac{k}{\sinh(2kh)}\sqrt{\frac{\nu\Omega_h}{2}},\qquad
\alpha_{N,h}\simeq\frac{\Gamma_{N,h}}{v_g^{(h)}}.}
\label{eq:newtonfinite}
\end{equation}
The approximation requires a thin bottom layer relative to both $h$ and the outer horizontal scale; the full Newtonian finite-depth dispersion relation must be used when the required scale separation is absent \cite{LeBlondMainardi1987}.

\section{Power-law bulk dissipation and nonlinear envelopes}
\label{sec:deepPL}
\subsection{Deep-water coefficient from the complete strain invariant}
Using Eq.~\eqref{eq:deepfield}, the independent strain components are
\begin{equation}
\begin{aligned}
D_{xx}&=-k\Omega Ae^{kz}\sin\theta,\quad D_{zz}=-D_{xx},\\
D_{xz}&=k\Omega Ae^{kz}\cos\theta.
\end{aligned}
\label{eq:tensor}
\end{equation}
The essential feature is that the \emph{equivalent} shear rate is independent of phase for this specific progressive deep-water potential field,
\begin{equation}
\dot\gamma=\sqrt{2\mathbf D:\mathbf D}=2k\Omega Ae^{kz}.
\label{eq:gammadot}
\end{equation}
The constitutive power density from Eq.~\eqref{eq:constitutive} is
\begin{equation}
\Phi_{\rm PL}=\bm\tau:\mathbf D
=K\dot\gamma^{n+1}.
\label{eq:powdensity}
\end{equation}
Thus the full-depth bulk power loss is determined explicitly, without fitting a uniform effective viscosity,
\begin{equation}
\boxed{\begin{aligned}
\Pw_{\rm PL}
&=K(2k\Omega A)^{n+1}\int_{-\infty}^0 e^{(n+1)kz}\dd z\\
&=\frac{K(2k\Omega)^{n+1}}{(n+1)k}A^{n+1}.
\end{aligned}}
\label{eq:powerdeep}
\end{equation}
Substituting $\Ew=\rho\Omega^2A^2/(2k)$ into Eq.~\eqref{eq:master}, we obtain the leading power-law bulk envelope equations
\begin{equation}
\boxed{\frac{\dd A}{\dd t}=-C_tA^n,\qquad
\frac{\dd A}{\dd x}=-C_xA^n,\qquad
C_x=\frac{C_t}{v_g},}
\label{eq:envelopeODE}
\end{equation}
where
\begin{equation}
\boxed{C_t=\frac{4Kk^2}{(n+1)\rho}
(2k\Omega)^{n-1}.}
\label{eq:Ct}
\end{equation}
At $n=1$, $K=\mu$, Eq.~\eqref{eq:Ct} recovers exactly $2\nu k^2$ within the weak-loss calculation. For fixed $k$ and a locally uniform medium, the temporal and spatial solutions are
\begin{equation}
\frac{A}{A_0}=\begin{cases}
\exp(-C_tt),&n=1,\\
\big[1+(n-1)C_ttA_0^{n-1}\big]^{1/(1-n)},&n\ne1,
\end{cases}
\label{eq:timedecay}
\end{equation}
with $C_tt$ replaced by $C_xx$ for spatial decay. The bracket must remain positive. Defining local logarithmic rates, rather than forcing a constant-exponential fit, gives
\begin{equation}
\boxed{\begin{aligned}
\gamma_t(A)&=-\frac{\dd\ln A}{\dd t}=C_tA^{n-1},\\
\alpha_{\rm loc}(A)&=-\frac{\dd\ln A}{\dd x}=\frac{C_tA^{n-1}}{v_g}.
\end{aligned}}
\label{eq:rates}
\end{equation}
The model predicts $\gamma_t\propto A^{n-1}$: a shear-thinning liquid ($0<n<1$) has a larger \emph{local} rate as amplitude decreases, whereas a shear-thickening one ($n>1$) has a smaller local rate as amplitude decreases. This does \emph{not} imply that a thinning liquid always damps faster than a thickening liquid at matched material conditions: comparison requires a common reference viscosity and shear rate. For $n>1$ the formal envelope has an algebraic tail; for $0<n<1$ it has a formal finite endpoint, but the growing $\gamma_t/\Omega$ and singular zero-shear apparent viscosity invalidate interpreting that endpoint as physical wave arrest.

\subsection{Wave-number signatures and constitutive matching}
The gravity and capillary limits of Eq.~\eqref{eq:Ct} give
\begin{equation}
\begin{aligned}
C_t&\propto\begin{cases}
k^{(3n+1)/2},&\text{gravity},\\
k^{(5n-1)/2},&\text{capillary},
\end{cases}\\
\alpha_{\rm loc}&\propto A^{n-1}\begin{cases}
k^{(3n+2)/2},&\text{gravity},\\
k^{(5n-2)/2},&\text{capillary}.
\end{cases}
\end{aligned}
\label{eq:exponents}
\end{equation}
For a comparable family of rheologies, specifying the same \emph{number} $K$ across different $n$ is inappropriate because the units of $K$ depend on $n$. Instead we require $\mu_{\rm app}(\dot\gamma_{\rm ref})=\mu_{\rm ref}$ at an independently stated positive reference shear rate:
\begin{equation}
K_n=\mu_{\rm ref}\dot\gamma_{\rm ref}^{1-n},\qquad
\frac{\gamma_{t,n}(A)}{2\mu_{\rm ref}k^2/\rho}
=\frac{2}{n+1}\left(\frac{2k\Omega A}{\dot\gamma_{\rm ref}}\right)^{n-1}.
\label{eq:matching}
\end{equation}
Thus the predicted ordering of rates depends jointly on $n$, $A$, $k$, and the chosen rheological reference condition.

\section{Finite-depth power-law bulk and bottom-layer losses}
\label{sec:finite}
\subsection{Phase-resolved bulk dissipation}
For $H=kh$ and $q=k(z+h)$, the standard finite-depth outer potential velocities are
\begin{equation}
\begin{aligned}
u&=\Omega_hA\frac{\cosh q}{\sinh H}\cos\theta,\\
w&=\Omega_hA\frac{\sinh q}{\sinh H}\sin\theta.
\end{aligned}
\label{eq:finitefield}
\end{equation}
They enforce impermeability $w(-h)=0$ but not the no-slip condition. Differentiation of Eq.~\eqref{eq:finitefield} gives
\begin{equation}
\dot\gamma=\frac{2k\Omega_hA}{\sinh H}
\left(\cosh^2q\sin^2\theta+\sinh^2q\cos^2\theta\right)^{1/2}.
\label{eq:finiteinvariant}
\end{equation}
In contrast with deep water, $\dot\gamma$ depends on phase; generally $\mean{\dot\gamma^{n+1}}\ne\mean{\dot\gamma}^{\,n+1}$. Retaining the nonlinear power \emph{inside} the phase average defines
\begin{widetext}
\begin{equation}
\boxed{J_n(H)=\frac{1}{2\pi}\int_0^{2\pi}\int_0^H
\left[\frac{\big(\cosh^2q\sin^2\theta+\sinh^2q\cos^2\theta\big)^{1/2}}{\sinh H}\right]^{n+1}
\dd q\,\dd\theta.}
\label{eq:J}
\end{equation}
\end{widetext}
Using the leading conservative energy per surface area $\Ew_h=\rho\Omega_h^2 A^2/(2k\tanh H)$, the integrated bulk power and amplitude coefficient become
\begin{equation}
\Pw_{{\rm bulk},h}=\frac Kk(2k\Omega_hA)^{n+1}J_n(H),
\label{eq:Pfinite}
\end{equation}
\begin{equation}
\boxed{\begin{aligned}
\frac{\dd A}{\dd t}&=-C_{t,h}^{\rm bulk}A^n,\\
C_{t,h}^{\rm bulk}&=\frac{4Kk^2}{\rho}(2k\Omega_h)^{n-1}
J_n(H)\tanh H.
\end{aligned}}
\label{eq:Cfinite}
\end{equation}
Two exact checks follow from the quadrature, independent of any fitted coefficient:
\begin{equation}
\boxed{J_1(H)=\tfrac12\coth H,\qquad
\lim_{H\to\infty}J_n(H)=\frac1{n+1}.}
\label{eq:Jncheck}
\end{equation}
Consequently $C_{t,h}^{\rm bulk}=2\nu k^2$ for $n=1$ and recovers Eq.~\eqref{eq:Ct} as $H\to\infty$. These are consistency checks of the \emph{bulk} construction; a real finite-depth no-slip experiment has additional dissipation.

\subsection{Nonlinear bottom-layer problem and conditional composite}
At the bed the exterior velocity has amplitude $U_b=\Omega_hA/\sinh H$. Let $y=z+h$ and write the local horizontal velocity as $u=u_e(t)+v(y,t)$, with $u_e=U_b\cos(\Omega_ht)$ and $v$ a wall-velocity \emph{defect}. In the locally unidirectional thin-layer approximation, subtracting the outer acceleration yields
\begin{equation}
\begin{aligned}
\rho\,\partial_tv&=\partial_y\left[K|\partial_yv|^{n-1}\partial_yv\right],\\
v(0,t)&=-u_e(t),\qquad v(\infty,t)=0.
\end{aligned}
\label{eq:wallPDE}
\end{equation}
This is an oscillatory power-law Stokes problem, not a Newtonian exponential profile with a redefined uniform viscosity \cite{PritchardMcArdleWilson2011}. The scaling choice
\begin{equation}
\begin{aligned}
\delta_n&=\left(\frac{2KU_b^{n-1}}{\rho\Omega_h}\right)^{1/(n+1)},\\
Y&=y/\delta_n,\quad T=\Omega_ht,\quad V=v/U_b.
\end{aligned}
\label{eq:delta}
\end{equation}
reduces Eq.~\eqref{eq:wallPDE} to
\begin{equation}
\begin{aligned}
\partial_TV&=\tfrac12\partial_Y[|V_Y|^{n-1}V_Y],\\
V(0,T)&=-\cos T,\qquad V(\infty,T)=0.
\end{aligned}
\label{eq:wallND}
\end{equation}
A sufficiently localized periodic solution defines the nonnegative dimensionless loss integral
\begin{equation}
I_n=\int_0^\infty\mean{|V_Y|^{n+1}}_T\,\dd Y,\quad
\Pw_b=\frac{KU_b^{n+1}}{\delta_n^n}I_n.
\label{eq:I}
\end{equation}
Its numerical value cannot be inferred from a constant-viscosity substitution. At $n=1$ the exact solution $V=-e^{-Y}\cos(T-Y)$ gives $I_1=1/2$ and the Newtonian bottom-layer term in Eq.~\eqref{eq:newtonfinite}. For $n\ne1$, the nonlinear periodic boundary-value problem is evaluated numerically below; the result is a coefficient of the reduced wall model, not a solution of the full free-surface problem.

\subsection{Numerical determination of the nonlinear wall-loss coefficient}
\label{subsec:wallnumerical}
To obtain $I_n$ rather than treat it as a fitting parameter, Eq.~\eqref{eq:wallND} is discretized by conservative centered shear fluxes on $0\leq Y\leq Y_{\max}$ and second-order backward differentiation (BDF2) in periodic time; the first step uses backward Euler. At each step, a damped Newton solve enforces the implicit nonlinear diffusion equation. We use the differentiable regularization $F_\varepsilon(s)=s(s^2+\varepsilon^2)^{(n-1)/2}$ for the discrete stress flux, then independently check the sensitivity of $I_n$ as $\varepsilon$ is reduced. The reported solutions use $Y_{\max}=16$, 960 spatial intervals, 768 steps per period, $\varepsilon=10^{-5}$, and periodic endpoint agreement below $7\times10^{-7}$ in normalized velocity. The numerical integral is evaluated using the unregularized dissipation $|V_Y|^{n+1}$; the regularized flux work is also checked separately.

As an energy check, multiplication of the wall equation by the defect $V$, integration in $Y$, and averaging over one period give
\begin{equation}
\boxed{I_n=\mean{\cos T\,|V_Y(0,T)|^{n-1}V_Y(0,T)}_T.}
\label{eq:wallworkI}
\end{equation}
Here the right side is the power supplied by the effective defect-flow boundary forcing, not work done by the physically stationary bottom. Discrete volume and boundary evaluations agree within $2\times10^{-7}$ relative error at the finest reported resolution. Representative values are
\begin{equation}
\boxed{I_{0.6}\simeq0.5432,\quad I_1=0.5000,\quad I_{1.4}\simeq0.4708.}
\label{eq:Innumerical}
\end{equation}
The finest-grid volume values are 0.543165, 0.499971, and 0.470768, respectively; rounding to four decimal places is consistent with the observed mesh/time refinement toward the Newtonian exact result. Domain and flux-regularization sensitivity were also evaluated, and the discretized $I_n$ shifts by less than approximately $5\times10^{-4}$ over the finest practical spatial/time refinements, with further convergence required for additional significant digits. These tests provide numerical evidence for the reduced-wall coefficient, not experimental validation of matched surface-wave damping. Oscillatory power-law Stokes layers and their non-Newtonian penetration behavior have been analyzed previously \cite{PritchardMcArdleWilson2011}; the present use of $I_n$ is in the wave-specific finite-depth dissipation budget.

The wall thickness scales with amplitude as $\delta_n\propto A^{(n-1)/(n+1)}$ at fixed $(k,h,K,n)$. Equation~\eqref{eq:I} therefore gives $\Pw_b\propto A^{(3n+1)/(n+1)}$ and a \emph{wall} logarithmic rate $\Gamma_b\propto A^{(n-1)/(n+1)}$, distinct from the bulk exponent $n-1$. With $\mathcal U_b=\Omega_h/\sinh H$ and $\mathcal D_n=[2K\mathcal U_b^{n-1}/(\rho\Omega_h)]^{1/(n+1)}$, define
\begin{equation}
C_{t,h}^{\rm wall}
=\frac{k\tanh H\,K\mathcal U_b^{n+1}I_n}
{\rho\Omega_h^2\mathcal D_n^n}.
\label{eq:Cwall}
\end{equation}
\emph{Only if} the outer bulk region and inner wall region remain asymptotically separated, their leading powers can be added to form the proposed composite amplitude laws
\begin{equation}
\boxed{\begin{aligned}
\frac{\dd A}{\dd t}&\simeq -C_{t,h}^{\rm bulk}A^n
-C_{t,h}^{\rm wall}A^{2n/(n+1)},\\
\frac{\dd A}{\dd x}&\simeq -\frac{C_{t,h}^{\rm bulk}A^n
+C_{t,h}^{\rm wall}A^{2n/(n+1)}}{v_g^{(h)}}.
\end{aligned}}
\label{eq:composite}
\end{equation}
The first power is supported by the evaluated outer-field quadrature; the wall term is conditional on solving Eq.~\eqref{eq:wallND} and validating matched asymptotics. It is not an exact superposition of two nonlinear velocity solutions. In particular, the algebraic tails of shear-thinning Stokes layers can invalidate simple separation despite a nominally small $\delta_n/h$.

\section{Gravity--capillary--damping regime coordinates}
\label{sec:regimemap}
\subsection{Restoring-force partition versus viscous loss}
The classical gravity--capillary crossover describes the relative size of the two restoring terms, not a transition between underdamped and overdamped motion. Define the dimensionless wavenumber $q_c=k\ell_c$, with $\ell_c=\sqrt{\sigma/(\rho g)}$. The capillary and gravitational fractions of the conservative restoring term are
\begin{equation}
\boxed{f_c(q_c)=\frac{q_c^2}{1+q_c^2},\qquad
f_g(q_c)=\frac{1}{1+q_c^2},\quad f_c+f_g=1.}
\label{eq:forcefractions}
\end{equation}
Equal restoring contributions occur at $q_c=1$. For descriptive plots we designate $q_c<1/3$ gravity-dominated ($f_g>0.9$), $1/3\leq q_c\leq3$ mixed, and $q_c>3$ capillary-dominated ($f_c>0.9$). These 10\%/90\% labels are \emph{conventions}, not singularities of the dispersion relation. Observational gravity--capillary partitions have appeared previously \cite{LaxagueEtal2018}; here they are combined with the constitutive damping predictions of Eq.~\eqref{eq:Ct}.

For comparison across different flow indices, use $\mu_{\rm ref}$ at $\dot\gamma_{\rm ref}$ as in Eq.~\eqref{eq:matching}. Introducing the local steepness $\epsilon=kA$, the dimensionless temporal damping rate becomes
\begin{equation}
\boxed{\zeta_n(q_c,\epsilon)=\frac{\Gamma_n(A,k)}{\Omega(k)}
=\frac{2}{n+1}\frac{2\nu_{\rm ref}k^2}{\Omega(k)}
\left(\frac{2\epsilon\Omega(k)}{\dot\gamma_{\rm ref}}\right)^{n-1},}
\label{eq:zetamap}
\end{equation}
where $k=q_c/\ell_c$ and $\nu_{\rm ref}=\mu_{\rm ref}/\rho$. The factor $2/(n+1)$ originates from the full-depth dissipation integral, not from an adjustable apparent-viscosity factor. An illustrative admissibility contour $\zeta_n=0.1$ separates displayed model points with relatively slow decay from a region where this perturbative prediction should not be trusted. This is \emph{not} an exact critical-damping or propagation boundary. Independent checks $\epsilon\ll1$ and $\alpha_{\rm loc}/k\ll1$ remain necessary.

\subsection{Local propagation distance in wavelengths}
For a prescribed local amplitude, the weak-loss plane-wave relation $\alpha_{\rm loc}=\Gamma_n/v_g$ and the conservative group speed yield
\begin{equation}
\frac{kv_g}{\Omega}=\frac{1+3q_c^2}{2(1+q_c^2)},\qquad
\boxed{\frac{\alpha_{\rm loc}}{k}=
\frac{2(1+q_c^2)}{1+3q_c^2}\zeta_n.}
\label{eq:alphadimmap}
\end{equation}
The associated \emph{local} amplitude attenuation length $L_{\rm loc}=1/\alpha_{\rm loc}$, divided by the carrier wavelength $\lambda=2\pi/k$, is
\begin{equation}
\boxed{N_\lambda=\frac{L_{\rm loc}}{\lambda}
=\frac{1+3q_c^2}{4\pi(1+q_c^2)\zeta_n}.}
\label{eq:Nwaves}
\end{equation}
For Newtonian waves $L_{\rm loc}$ is the e-folding length of the exponential envelope. For $n\ne1$, it is a \emph{local differential} length: $\alpha_{\rm loc}$ changes with $A$, so $N_\lambda$ is not the total number of wavelengths a wave will travel before an arbitrary detection threshold. Instrument noise, excitation geometry, reflections, and radial spreading are not encoded in this measure.

\subsection{Calibration-dependent equal-damping contour}
The ratio of the bulk power-law rate to the reference Newtonian rate is Eq.~\eqref{eq:matching}. Their equality, for $n\ne1$, defines
\begin{equation}
\boxed{S_*(n)=\left(\frac{n+1}{2}\right)^{1/(n-1)},\qquad
\epsilon_*(q_c;n)=\frac{\dot\gamma_{\rm ref} S_*(n)}{2\Omega(q_c)}.}
\label{eq:equalmap}
\end{equation}
Here $S=2\epsilon\Omega/\dot\gamma_{\rm ref}$ is the wave-generated shear scale relative to the arbitrary but explicitly declared viscosity-matching shear rate. Equation~\eqref{eq:equalmap} is an \emph{equal-rate locus between calibrated models}, not a new rheological material phase boundary. The locus is physically informative only where it satisfies the weak-loss and small-steepness conditions simultaneously.

\subsection{Fixed-steepness scaling and a conditional damping maximum}
The wavenumber trends depend on how amplitude is held during a scan. At fixed physical $A$ the powers in Eq.~\eqref{eq:exponents} apply; at \emph{fixed steepness} $\epsilon=kA$ instead, Eq.~\eqref{eq:zetamap} yields
\begin{equation}
\zeta_n\propto k^2\Omega^{n-2}\propto
\begin{cases}
k^{(n+2)/2},&q_c\ll1,\\
k^{(3n-2)/2},&q_c\gg1.
\end{cases}
\label{eq:zetaasymp}
\end{equation}
Using $\Omega^2=(g/\ell_c)q_c(1+q_c^2)$ and differentiating at fixed $\epsilon$,
\begin{equation}
\frac{\dd\ln\zeta_n}{\dd\ln q_c}
=\frac{(n+2)+(3n-2)q_c^2}{2(1+q_c^2)}.
\label{eq:zetaslope}
\end{equation}
For $0<n<2/3$ this derivative vanishes at
\begin{equation}
\boxed{q_{c,\mathrm{peak}}=\sqrt{\frac{n+2}{2-3n}}\qquad(0<n<2/3).}
\label{eq:zetapeak}
\end{equation}
For example, $n=0.6$ gives $q_{c,\mathrm{peak}}=\sqrt{13}\simeq3.61$. This is a prediction for \emph{normalized temporal damping} $\Gamma/\Omega$ under a fixed-steepness protocol, not necessarily a maximum of the dimensional spatial attenuation $\alpha$, and is not credible if the predicted maximum falls outside the constitutive or weak-loss range.

\subsection{Finite-depth extension and a third mechanism coordinate}
The restoring-force fraction in Eq.~\eqref{eq:forcefractions} remains the same for the conservative finite-depth law because both terms share $\tanh(kh)$; however $\Omega_h$, $v_g^{(h)}$, and $J_n(kh)$ change the attenuation contours. Accordingly, a finite-depth regime description requires $q_c$, $H=kh$, and $\epsilon$, together with the rheological parameters. A useful \emph{candidate} third mechanism coordinate for a no-slip bottom is
\begin{equation}
\boxed{R_{bw}=\frac{\Gamma_b}{\Gamma_{\rm bulk}}
=\frac{C_{t,h}^{\rm wall}}{C_{t,h}^{\rm bulk}}
A^{-n(n-1)/(n+1)}.}
\label{eq:regimeRbw}
\end{equation}
The contour $R_{bw}=1$ identifies equal wall and distributed bulk loss. For Newtonian flow the wall coefficient is known analytically. For $n=0.6$ and $n=1.4$, the numerically evaluated $I_n$ in Eq.~\eqref{eq:Innumerical} also specifies the \emph{reduced-model} absolute ratio at chosen $K$, $A$, $k$, and $h$. Its location is thus now calculable, although validity of adding the two dissipative regions remains conditional on asymptotic separation. At fixed $(k,h,K,n)$ the relative amplitude drift follows independently of the coefficient:
\begin{equation}
\boxed{\begin{aligned}
\mathcal D_n(\chi_A)&\equiv
\frac{R_{bw}(A)}{R_{bw}(A_{\rm ref})}
=\chi_A^{-n(n-1)/(n+1)},\\
\chi_A&\equiv\frac{A}{A_{\rm ref}}.
\end{aligned}}
\label{eq:drift}
\end{equation}
For decreasing $A$, the model predicts a relative shift toward bulk-dominated loss for $0<n<1$ and toward bottom-layer loss for $n>1$. The corresponding absolute equality curve depends on material calibration, wave steepness, depth, and the computed $I_n$. Neither the computed wall coefficient nor the resulting map independently validates the coupled free-surface model.

\section{Results: analytical predictions and diagnostic consequences}
\label{sec:results}
The following results are consequences of Eqs.~\eqref{eq:envelopeODE}--\eqref{eq:composite}, not measurements or validations of a generalized-Newtonian free-surface simulation. The plotted curves were generated directly from the displayed analytical expressions. We use $n=0.6$, $1$, and $1.4$ as illustrative indices and state the normalization independently in each caption. Physical interpretation is limited to parameter and amplitude ranges satisfying the applicability conditions of Sec.~\ref{sec:validity}.

\subsection{The constitutive exponent changes the shape of the envelope}
\begin{figure}[tbp]
\includegraphics[width=\columnwidth]{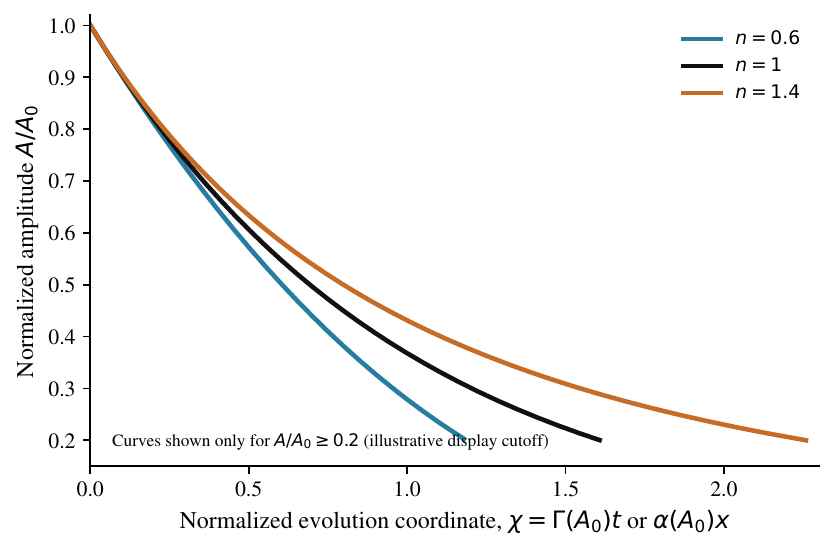}
\caption{\label{fig:envelopes}Theory-only normalized bulk envelopes from Eq.~\eqref{eq:scaledenv}. Each curve is matched to the same initial logarithmic decay coefficient. The illustrative plotting cutoff $A/A_0\ge0.2$ does not establish a physical range of validity; actual weak-loss conditions must be checked case by case.}
\end{figure}

At a common initial local logarithmic rate $\Gamma(A_0)=C_tA_0^{n-1}$, Eq.~\eqref{eq:timedecay} gives the dimensionless family
\begin{equation}
\frac{A}{A_0}=
\begin{cases}
\exp(-\chi),&n=1,\\
[1+(n-1)\chi]^{1/(1-n)},&n\ne1,
\end{cases}
\quad
\chi=\Gamma(A_0)t
\label{eq:scaledenv}
\end{equation}
(or $\chi=\alpha(A_0)x$ for plane spatial decay). Figure~\ref{fig:envelopes} shows that a shear-thinning model decreases more steeply as its amplitude falls, whereas a shear-thickening model develops a slower late-stage reduction in its local logarithmic rate. The curve cutoff is for illustration, not a universal physical threshold.

\subsection{Finite-depth dissipation is rheology dependent}
\begin{figure}[tbp]
\includegraphics[width=\columnwidth]{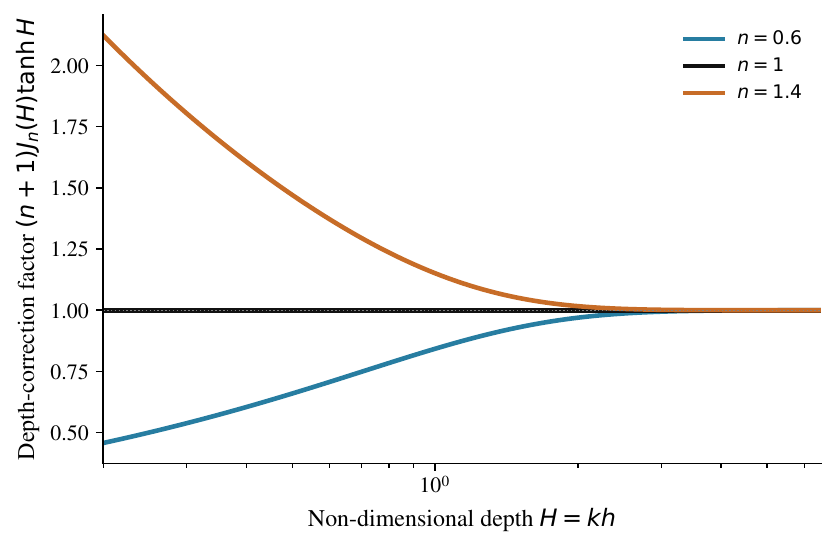}
\caption{\label{fig:depthfactor}Theory-only factor $Q_n=(n+1)J_n(H)\tanh H$ evaluated by numerical quadrature of Eq.~\eqref{eq:J}. The Newtonian curve is identically one; all models approach unity at large $H=kh$. This is a distributed \emph{bulk} correction, not total no-slip attenuation.}
\end{figure}

A useful purely geometrical measure of the finite-depth bulk correction is
\begin{equation}
Q_n(H)=(n+1)J_n(H)\tanh H.
\label{eq:Q}
\end{equation}
From Eq.~\eqref{eq:Jncheck}, $Q_1(H)=1$ for every depth and $Q_n\to1$ as $H\to\infty$. For $n\ne1$, $Q_n$ need not equal unity. Numerical quadrature of the exact integral in Eq.~\eqref{eq:J} gives, for example, $Q_{0.6}(1)\simeq0.843$ and $Q_{1.4}(1)\simeq1.151$; these numbers are analytical-model evaluations, not experimental results. Figure~\ref{fig:depthfactor} isolates the influence of the finite-depth strain distribution from the separate influence of $\Omega_h$ on $C_{t,h}^{\rm bulk}$. In particular, the cancellation that produces depth-independent \emph{bulk} damping at $n=1$ is not generally inherited by nonlinear constitutive laws.

\subsection{Bulk and bottom attenuation have distinct amplitude signatures}
\begin{figure}[tbp]
\includegraphics[width=\columnwidth]{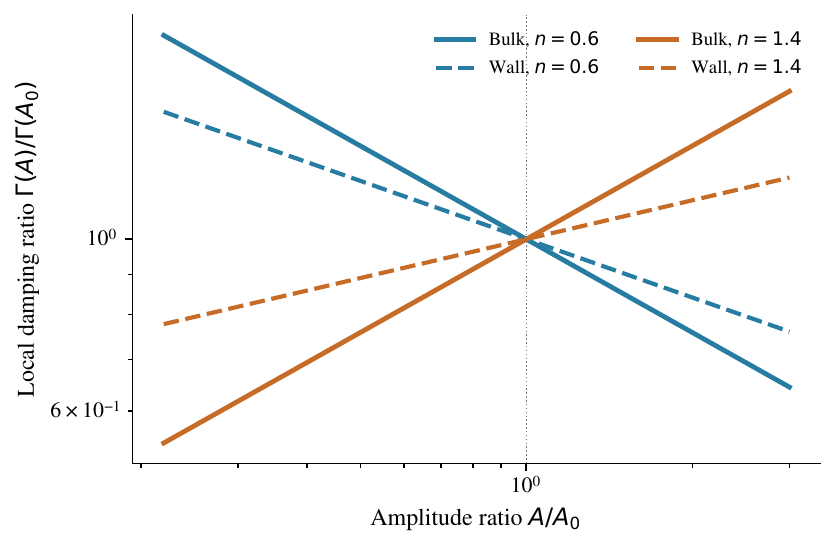}
\caption{\label{fig:slopes}Predicted normalized local attenuation powers $\Gamma(A)/\Gamma(A_0)$ for the bulk (solid) and wall (dashed) mechanisms. Each individual mechanism is separately normalized to one at $A=A_0$; relative \emph{magnitudes} require both the numerically evaluated $I_n$ and specified $(k,h,A,K,n)$; normalized curves alone do not reveal which region dominates.}
\end{figure}

Write $\Gamma_{\rm bulk}=B A^{n-1}$ and $\Gamma_b=W A^{(n-1)/(n+1)}$, with amplitude-independent $B,W$ at fixed $(k,h,n,K)$. Their ratio satisfies
\begin{equation}
\boxed{R(A)=\frac{\Gamma_b}{\Gamma_{\rm bulk}}
=\frac WB A^{-n(n-1)/(n+1)}.}
\label{eq:ratio}
\end{equation}
Figure~\ref{fig:slopes} illustrates the distinct slopes without choosing an unsupported non-Newtonian value for $I_n$. If both mechanisms contribute, their total logarithmic slope is a weighted average,
\begin{equation}
\frac{\dd\ln\Gamma_{\rm total}}{\dd\ln A}
=\frac{(n-1)\Gamma_{\rm bulk}
+\frac{n-1}{n+1}\Gamma_b}
{\Gamma_{\rm bulk}+\Gamma_b}.
\label{eq:mixslope}
\end{equation}
Consequently an intermediate measured slope cannot be interpreted as $n-1$ without excluding or estimating wall friction. This is a potential identifiability limitation and a mechanism-selection prediction, rather than a measured crossover.

\begin{figure}[tbp]
\includegraphics[width=\columnwidth]{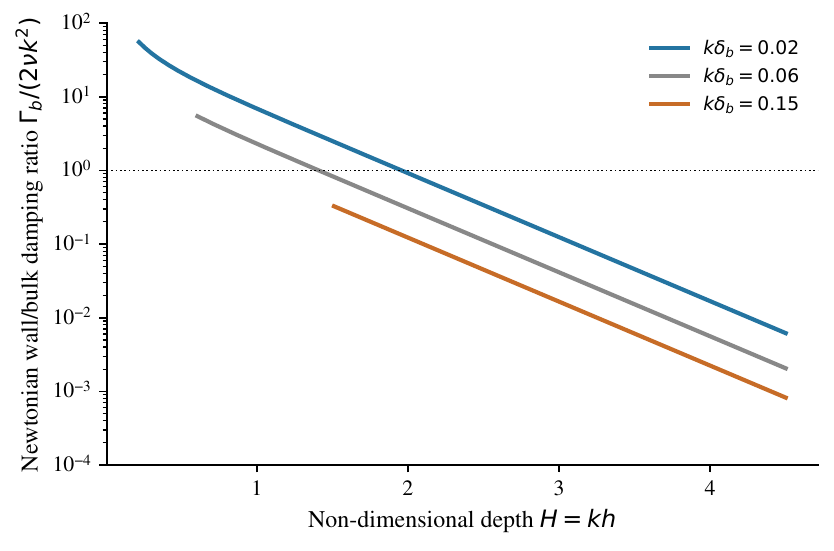}
\caption{\label{fig:wallratio}Newtonian thin-bottom-layer ratio calculated from Eq.~\eqref{eq:wallratioN} for prescribed $k\delta_b$ values. Displayed curves use the illustrative mask $\delta_b/h=(k\delta_b)/(kh)<0.1$; weak total damping must also be verified.}
\end{figure}

The importance of rigid-bottom damping can already be illustrated quantitatively in the \emph{Newtonian} limit. Inserting $\delta_b=\sqrt{2\nu/\Omega_h}$ into Eq.~\eqref{eq:newtonfinite} gives
\begin{equation}
\frac{\Gamma_b}{\Gamma_{\rm bulk}}=\frac{1}{2k\delta_b\sinh(2kh)}.
\label{eq:wallratioN}
\end{equation}
For sufficiently shallow depths the bed contribution can exceed bulk dissipation even while $k\delta_b$ is small (Fig.~\ref{fig:wallratio}), provided $\delta_b\ll h$ also holds. The figure is a Newtonian asymptotic diagnostic, not evidence that the nonlinear composite has been validated.

\subsection{Restoring-force partition and the weak-loss window}
\begin{figure}[tbp]
\centering
\includegraphics[width=\columnwidth]{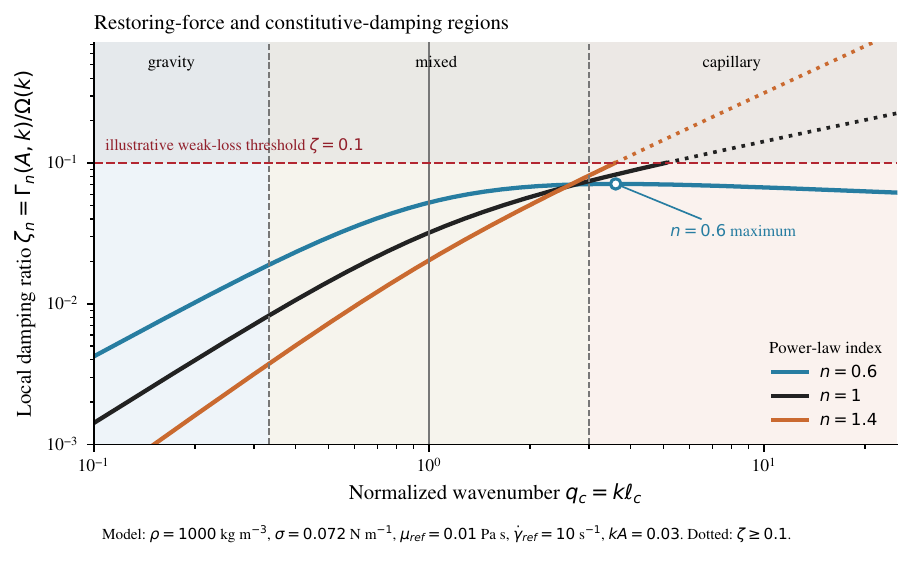}
\caption{\label{fig:gc_damping_map}Theory-only gravity--capillary--damping map for $\rho=1000$ kg\,m$^{-3}$, $\sigma=0.072$ N\,m$^{-1}$, $\mu_{\rm ref}=0.01$ Pa\,s at $\dot\gamma_{\rm ref}=10$ s$^{-1}$ and fixed steepness $kA=0.03$. The force-transition boundaries follow Eq.~\eqref{eq:forcefractions}; the curves evaluate Eq.~\eqref{eq:zetamap}. Dotted segments violate the illustrative $\zeta<0.1$ criterion. The marked $n=0.6$ maximum is the stationary point of $\Gamma/\Omega$ at fixed steepness, not a peak of $\Gamma$ or $\alpha$. Neither the band edges nor the horizontal threshold is a physical bifurcation.}
\end{figure}

The physically different coordinates of Eqs.~\eqref{eq:forcefractions} and \eqref{eq:zetamap} are placed on the same diagram in Fig.~\ref{fig:gc_damping_map}. The vertical lines $q_c=1/3$, 1, and 3 delimit a chosen 10\%--90\% transition band and equal restoring contributions. In contrast, the three attenuation curves move with the calibrated power-law index at the same steepness $\epsilon=0.03$. The horizontal line $\zeta=0.1$ is a stated \emph{model-screening tolerance}; dotted curve segments cannot be interpreted by the asymptotic model without additional analysis. For the illustrative parameter set, the $n=0.6$ curve reaches a local maximum near $q_c=3.61$, marked directly in the figure; the maximum lies on the capillary side of restoring-force equality and follows Eq.~\eqref{eq:zetapeak}. The crossover of forcing contributions does not imply a corresponding universal crossover of constitutive damping.

\subsection{Local attenuation length and observation conditions}
\begin{figure*}[tbp]
\centering
\includegraphics[width=.95\textwidth]{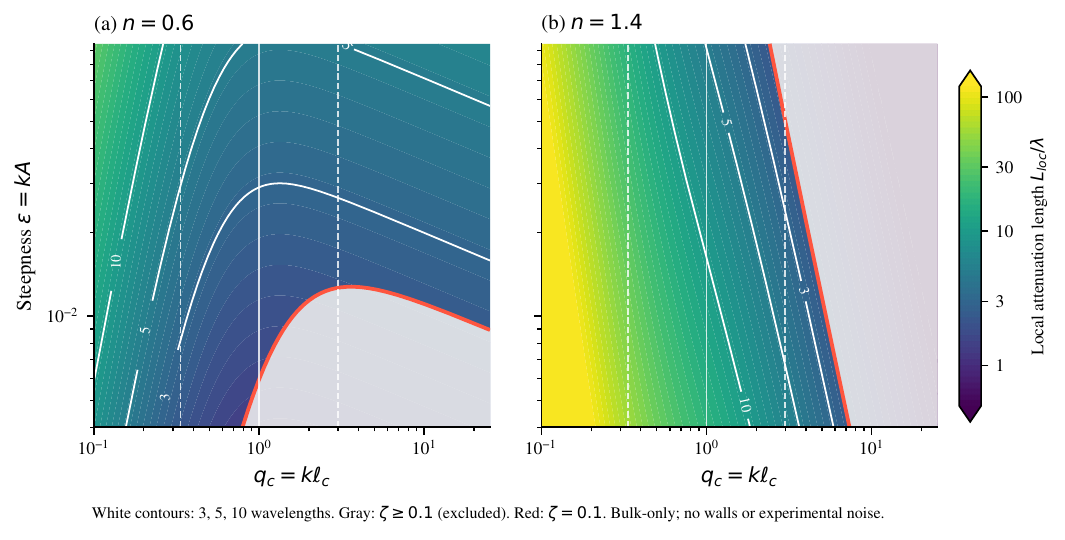}
\caption{\label{fig:gc_observation_map}Bulk-only local observation maps of $N_\lambda=1/(\alpha_{\rm loc}\lambda)$ calculated from Eq.~\eqref{eq:Nwaves}, at the same $\rho$, $\sigma$, $\mu_{\rm ref}$, and $\dot\gamma_{\rm ref}$ as Fig.~\ref{fig:gc_damping_map}. Panels show shear-thinning ($n=0.6$) and shear-thickening ($n=1.4$) models. White contours label 3, 5, and 10 wavelengths, and the red contour denotes $\zeta=0.1$. Gray regions are masked as outside the selected weak-loss domain. This is a \emph{local} attenuation length, not an integrated travel distance or an instrument-level observability guarantee.}
\end{figure*}

Figure~\ref{fig:gc_observation_map} uses Eq.~\eqref{eq:Nwaves} to show how both wavelength and initial steepness change the local distance associated with an amplitude e-folding. The contours $N_\lambda=3$, 5, and 10 are illustrative choices for selecting a measurement window, not guaranteed signal-to-noise thresholds. The maps screen out $\zeta\geq0.1$, while remaining inside the displayed $\epsilon<0.1$ range. Because $\zeta_n\propto\epsilon^{n-1}$ at fixed $q_c$, reducing steepness \emph{increases} relative damping for the ideal shear-thinning model but \emph{decreases} it for the shear-thickening model. This conclusion concerns the local energy-model rate and should not be extrapolated into the zero-shear singularity.

\subsection{A rheological equality boundary distinct from $q_c=1$}
\begin{figure}[tbp]
\centering
\includegraphics[width=\columnwidth]{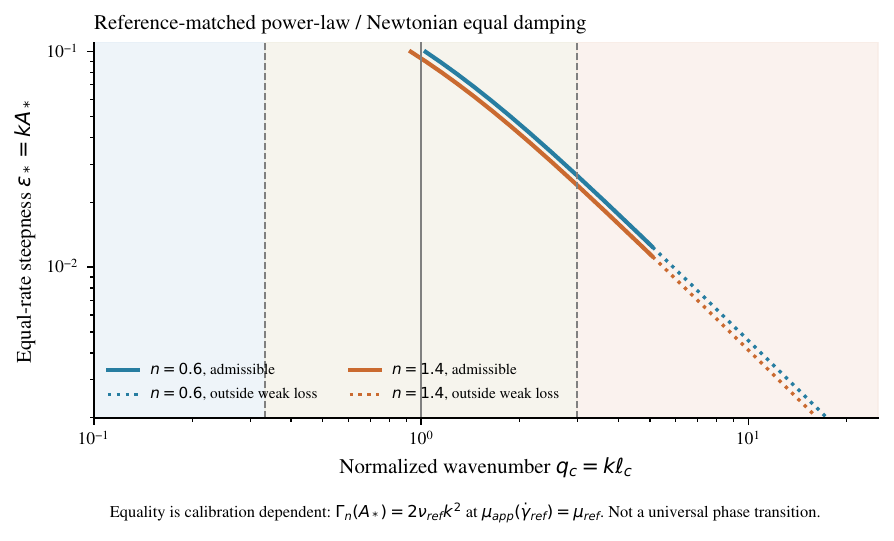}
\caption{\label{fig:gc_equal_map}Model-generated steepness at which each power-law rate equals the rate of a Newtonian fluid matched at $\mu_{\rm ref}$ and $\dot\gamma_{\rm ref}$, according to Eq.~\eqref{eq:equalmap}. Solid loci meet $\zeta<0.1$; dotted loci do not. This calibration-dependent equality is distinct from gravity--capillary restoring-force equality and is not a universal material transition.}
\end{figure}

The model permits a second, explicitly calibration-dependent line: equal damping of each power-law model and the viscosity-matched Newtonian fluid. Figure~\ref{fig:gc_equal_map} evaluates Eq.~\eqref{eq:equalmap}. Its location changes with the selected $\dot\gamma_{\rm ref}$ and differs qualitatively from the $q_c=1$ restoring-force equality, which is independent of the rheological index under the adopted conservative dispersion. In particular, an equality point that violates either $\epsilon\ll1$ or $\zeta\ll1$ must be discarded rather than labeled a physical change of damping mechanism.

\subsection{Finite-depth bulk--wall dominance with calculated nonlinear coefficients}
\begin{figure*}[t]
\centering
\includegraphics[width=0.98\textwidth]{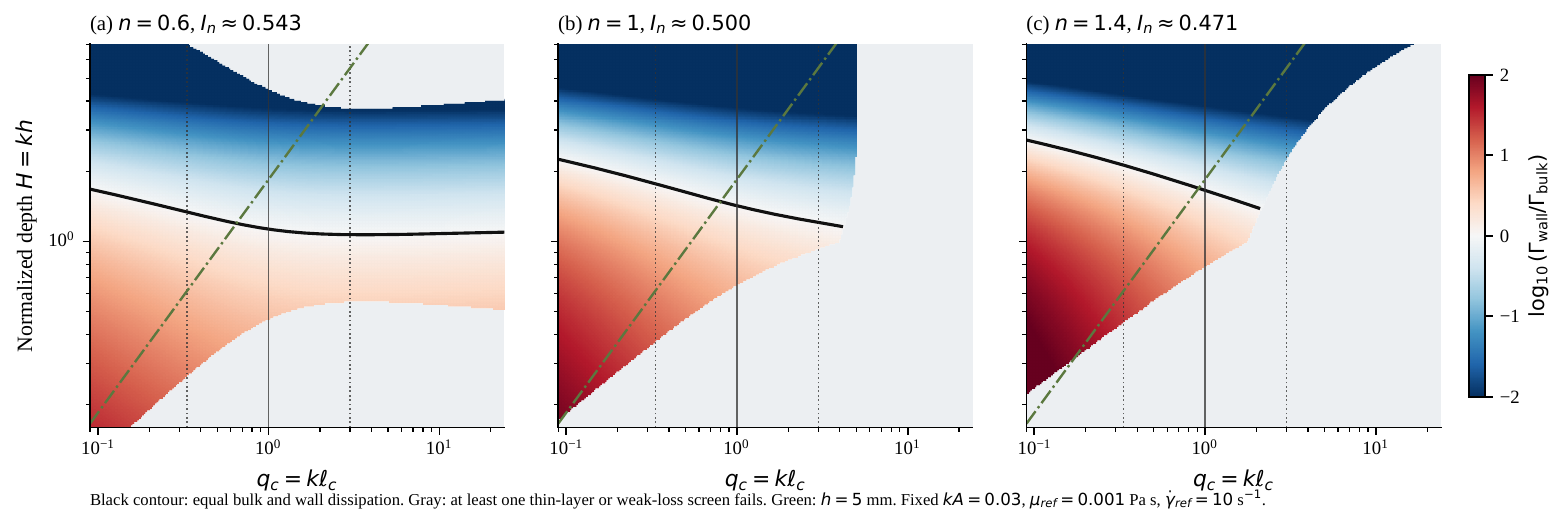}
\caption{\label{fig:bulk_wall_map}\textbf{Calculated finite-depth bulk--wall dominance at matched reference rheology.} Three panels show the ratio $\log_{10}(\Gamma_{\rm wall}/\Gamma_{\rm bulk})$ for (a) $n=0.6$, (b) $n=1$ (classical benchmark), and (c) $n=1.4$. Each power-law panel uses the periodic reduced-Stokes-layer integral $I_n$ from Eq.~\eqref{eq:Innumerical}; bulk dissipation is calculated from the independent finite-depth quadrature $J_n(kh)$. The horizontal coordinate $q_c=k\ell_c$ measures the relative capillary restoring force, whereas $H=kh$ measures depth; these are independent map axes. Warm colors indicate dominant wall dissipation, cool colors dominant distributed bulk dissipation, and the black contour marks their equality. The vertical solid line is the conservative gravity--capillary crossover $q_c=1$; dotted lines at $1/3$ and 3 bound a descriptive mixed-force band. The green dash--dot line represents a fixed depth $h=5$ mm. Gray regions fail one or more illustrative screens: $k\delta_n<0.1$, $\delta_n/h<0.1$, $\Gamma_{\rm total}/\Omega_h<0.1$, and $\Gamma_{\rm total}/(kv_g^{(h)})<0.1$. Parameters: $\rho=1000$ kg\,m$^{-3}$, $\sigma=0.072$ N\,m$^{-1}$, $g=9.81$ m\,s$^{-2}$, $kA=0.03$, and $K_n=\mu_{\rm ref}\dot\gamma_{\rm ref}^{1-n}$ with $\mu_{\rm ref}=0.001$ Pa\,s and $\dot\gamma_{\rm ref}=10$ s$^{-1}$. These regime contours depend on the matched-asymptotic bulk--wall hypothesis; their physical accuracy requires an independent full-flow benchmark.}
\end{figure*}

The two-dimensional regime coordinate $R_{bw}=\Gamma_{\rm wall}/\Gamma_{\rm bulk}$ can now be evaluated in the reduced composite model using Eq.~\eqref{eq:Innumerical}. The Newtonian case provides a stronger test because the analytic Stokes result is independently known:
\begin{equation}
\boxed{R_{bw,N}=\frac{1}{2\beta_N\sinh(2H)},\quad
\beta_N=k\sqrt{\frac{2\nu}{\Omega_h}}.}
\label{eq:RNewtonMap}
\end{equation}
For $n\ne1$, we compute $\mathcal P_{\rm bulk}$ from the full phase-resolved quadrature $J_n(H)$ and $\mathcal P_b=KU_b^{n+1}I_n/\delta_n^n$, so that $R_{bw}=\mathcal P_b/\mathcal P_{\rm bulk}$. Figure~\ref{fig:bulk_wall_map} compares the resulting dissipation maps for $n=0.6$, 1, and 1.4. All three models are matched to $\mu_{\rm app}(\dot\gamma_{\rm ref})=10^{-3}$ Pa\,s at $\dot\gamma_{\rm ref}=10$ s$^{-1}$ and are plotted at a common local steepness $kA=0.03$; this calibration avoids equating dimensionally different consistency values $K$. A fixed 5-mm depth appears as a green diagonal trajectory $H=hq_c/\ell_c$.

Warm and cool regions in Fig.~\ref{fig:bulk_wall_map} correspond to wall- and bulk-dominated dissipation, respectively. The black $R_{bw}=1$ contour lies generally away from the classical restoring-force line $q_c=1$: equal gravitational and capillary restoring terms do not imply equal viscous loss mechanisms. The detailed shape and admissible portion of the equality contour depend on $n$, the wave amplitude protocol, and the full nonlinear wall coefficient. Gray cells fail at least one chosen thin-layer or weak-loss criterion, and conclusions are limited to the unmasked part of the maps. This is a quantitative \emph{prediction of the reduced energy model}, not yet an independently verified surface-wave attenuation regime diagram.

\subsection{Numerical structure of the nonlinear oscillatory wall layer}
\begin{figure*}[t]
\centering
\includegraphics[width=0.97\textwidth]{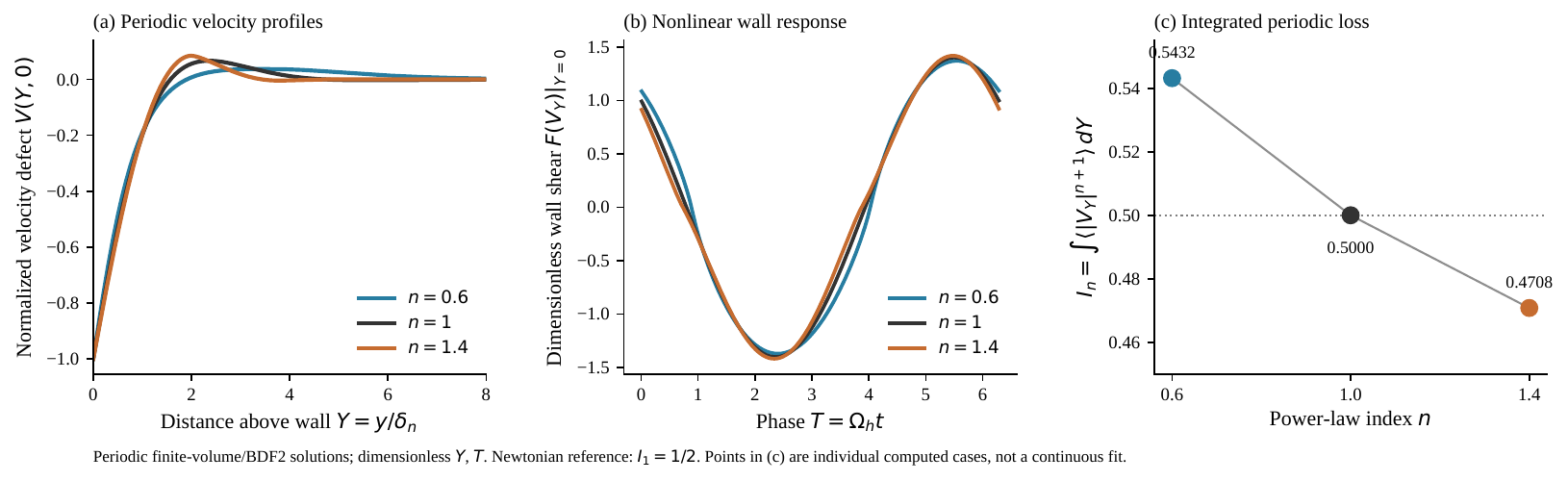}
\caption{\label{fig:wallnumerics}\textbf{Numerical periodic solutions of the reduced power-law Stokes layer.} (a) Normalized wall-velocity defect $V(Y,T)$ at phase $T=0$ for $n=0.6,1,1.4$; the wall condition is $V(0,0)=-1$, and the far-field defect approaches zero. (b) Dimensionless wall constitutive flux $F(V_Y)=V_Y(V_Y^2+\varepsilon^2)^{(n-1)/2}$ over one forcing period, with $\varepsilon=10^{-5}$. The response is not assumed sinusoidal at $n\ne1$. (c) Period-averaged integrated dissipation $I_n$ evaluated from the computed profiles. The Newtonian point is independently fixed by $I_1=1/2$; straight line segments between the three computed points are visual guides, not a fitted constitutive correlation. The results use a conservative spatial discretization and an implicit second-order BDF time integrator, with 960 intervals over $0\le Y\le16$ and 768 time steps per period. Periodic convergence, mesh refinement, far-boundary variation, and the wall-work identity were checked. These are reduced-boundary-layer results, not measured dissipation.}
\end{figure*}

Figure~\ref{fig:wallnumerics} shows why $I_n$ cannot be replaced by the Newtonian value at $n\ne1$. Although all cases share the same normalized wall forcing, the instantaneous velocity-defect profile and wall shear history are rheology-dependent. The shear-thinning response penetrates farther in the normalized coordinate than the shear-thickening response; the latter decays more sharply away from the wall, consistent with the qualitative Stokes-layer behavior established in Ref.~\cite{PritchardMcArdleWilson2011}. Integrating the nonlinear shearing cost produces $I_{0.6}>I_1>I_{1.4}$ for the three tested cases. These dimensionless inequalities by themselves do not imply the same ordering of dimensional wall attenuation because $K$, $U_b$, and $\delta_n$ also depend on rheology.

\subsection{Observable decay and rheological information}
For a bulk-dominated wave, the predicted local slope and prefactor give the formal inverse relations
\begin{equation}
\boxed{n=1+\frac{\dd\ln\gamma_t}{\dd\ln A},\qquad
K=\frac{(n+1)\rho\,\gamma_t(A)}{4k^2(2k\Omega A)^{n-1}}.}
\label{eq:inverse}
\end{equation}
The same local rate can be inferred from the spatial envelope through $\alpha_{\rm loc}=\gamma_t/v_g$, but measurements of temporal and spatial attenuation at the \emph{same} $(k,A)$ provide a consistency check, not automatically two independent rheological constraints. A credible inversion requires multiple independently calibrated amplitudes and/or wavenumbers, with density, depth, surface tension, source geometry, and bottom friction controlled. At non-negligible wall loss, Eq.~\eqref{eq:mixslope} must replace the bulk-only slope test.

For a localized, radially spreading surface wave the leading stationary energy flux across radius $r$ is $2\pi r v_g\Ew$. The radial envelope therefore satisfies, in the same leading local approximation,
\begin{equation}
\frac{\dd A}{\dd r}=-\frac{A}{2r}-\frac{C_t}{v_g}A^n.
\label{eq:radial}
\end{equation}
At $n=1$ this gives $A\propto r^{-1/2}e^{-\alpha_Nr}$. For $n\ne1$, geometrical spreading also changes the local constitutive dissipation, so multiplying a plane power-law envelope by $r^{-1/2}$ is generally incorrect. The radial equation is a conditional extension intended for experimental analysis, not a verified treatment of near-field excitation.

\section{Validity, comparison with literature, and next steps}
\label{sec:validity}
The predicted equations separate established reference physics from model-dependent constitutive predictions. In particular, Newtonian limits validate the normalization and leading weak-loss asymptote but do not show that a nonlinear free-surface flow actually follows the assumed potential field. Pritchard \emph{et al.} provide a relevant benchmark for the qualitative wall response, while Calvo \emph{et al.} and Hu \emph{et al.} demonstrate the importance of nonlinear rheology in \emph{different} wave regimes and geometries \cite{PritchardMcArdleWilson2011,CalvoEtal2026,HuChenLi2026}. The model thus contributes explicit predictions to compare against those studies, rather than claiming to supersede them.

The regime maps introduced in Sec.~\ref{sec:regimemap}, especially the three reduced-model panels of Fig.~\ref{fig:bulk_wall_map}, require a clear distinction between a physical restoring-force crossover ($q_c=1$), a conventional descriptive force-partition band ($1/3$--$3$), a chosen weak-damping plotting mask ($\zeta<0.1$), and an equal-attenuation curve defined relative to a selected $\dot\gamma_{\rm ref}$. No one of the latter three is a measured critical boundary. A useful local validity screen is $kA\ll1$, $\gamma_t(A)/\Omega\ll1$, and $\alpha_{\rm loc}/k\ll1$, where $\Omega$ is replaced by $\Omega_h$ at finite depth. One may use $0.1$ as a conservative plotting tolerance for some of these ratios, but no number is a universal critical-damping boundary. For the finite-depth composite, $k\delta_n\ll1$, $\delta_n/h\ll1$, and localization of the actual nonlinear wall shear are additionally required. The numerical power-law maps include $\delta_n$, the computed $I_n$, and total predicted damping, but they cannot by themselves certify asymptotic separation from the bulk or agreement with the full nonlinear free-surface problem. If the nominal wall tail overlaps a substantial fraction of the depth, the bulk and wall power cannot be reliably added. Source-region forcing, lateral walls, menisci, surface contamination, viscoelastic stresses, and temperature-dependent properties are outside the model.

The unregularized pure power law also requires special caution near small wave amplitude. For $n<1$, $\mu_{\rm app}=K\dot\gamma^{n-1}$ diverges as $\dot\gamma\to0$, and the predicted $\gamma_t(A)\sim A^{n-1}$ grows as $A$ decreases. Hence the formal finite-time zero from Eq.~\eqref{eq:timedecay} cannot be described as physical arrest. For $n>1$, the apparent viscosity tends to zero at vanishing strain, but effects neglected by the model can become comparatively important. Finite zero-shear viscosity and elastic memory require a regularized or viscoelastic constitutive model, not simply a change of $n$.

The immediate verification sequence is to (i) benchmark Newtonian deep- and finite-depth damping against the full linear viscous dispersion problem; (ii) check $J_1(H)$ and the deep-water limit of $J_n$ numerically; (iii) extend the reported nonlinear periodic wall-layer convergence checks and compare with independent published or numerical profiles; (iv) test the composite attenuation against an independently solved generalized-Newtonian free-surface/no-slip model; and (v) perform uncertainty-aware forward and inverse comparisons with calibrated liquids. A validation data set must not be synthesized from the same reduced ODE used to fit it and then presented as independent predictive evidence. The quantitative non-Newtonian equality contours obtained using the calculated $I_n$ are predictions of the reduced composite model, not evidence that the full free-surface dynamics obey this asymptotic separation.

\emph{Future interfacial extension.} The same mechanical-energy accounting can motivate a subsequent investigation of gravity--capillary waves at the interface between two immiscible liquids. That problem is \emph{not} solved here: both fluid layers carry inertia and bulk dissipation, the interface requires coupled kinematic and stress conditions, and an interfacial shear adjustment may contribute separate loss. A physically justified formulation must resolve the energy distribution and, where needed, matched interfacial layers rather than replacing the two constituent viscosities by an arbitrary weighted mean. In that subsequent work, an appropriate density-contrast capillary length will replace $\ell_c$ while separate layer-specific and interfacial energy participation measures enter the damping coordinate; the present $N_\lambda$ and $R_{bw}$ definitions should not be transferred unchanged without rederivation. Existing Newtonian layered-wave theory provides the necessary reference \cite{Dore1969,TroyKoseff2006,SuswanthRajan2025}. Thus the present free-surface calculations are a deliberately restricted first stage, not a claim of an already completed two-fluid model.

\section{Conclusions}
\label{sec:conclusions}
We have formulated a leading-order mechanical-energy model for damping of small-steepness gravity--capillary waves in memoryless power-law liquids. Classical Newtonian damping and group-velocity transport provide reference limits. Computing constitutive dissipation from the complete potential-wave strain tensor gives an explicit deep-water bulk coefficient and nonlinear amplitude equations in time and distance. A finite-depth quadrature retains the phase-dependent strain invariant and predicts a depth correction absent from the corresponding Newtonian \emph{bulk} rate. For a no-slip bottom, a separate power-law Stokes problem defines an additional, conditional energy loss with an amplitude exponent different from that of the bulk. The reduced periodic wall problem has now been solved numerically for $n=0.6$ and $1.4$, returning $I_n\simeq0.5432$ and $0.4708$, respectively, while the Newtonian solution yields $I_1=1/2$. The coexistence of bulk and wall loss can change the effective log-slope of attenuation and complicate inference of $n$ and $K$.

A combined regime description makes the physical separation explicit: $q_c=k\ell_c$ and $q_c=1$ classify conservative restoring forces; $\zeta_n$ and $N_\lambda$ describe rheology- and amplitude-dependent local attenuation and a candidate observation window; and $R_{bw}$ is a finite-depth mechanism ratio now evaluated for the three tested indices using the reduced Stokes-layer coefficients. Figure~\ref{fig:bulk_wall_map} exposes mechanism crossovers distinct from $q_c=1$, with gray masks excluding illustrative regions of weak-loss or thin-layer breakdown. The model also predicts a capillary-side maximum in the normalized temporal damping for $0<n<2/3$ at fixed steepness, subject to weak-loss validity. None of these threshold conventions constitutes a universal new phase transition.

The analytical Newtonian reductions, the finite-depth quadrature checks, and the displayed model-generated figures establish internal consistency, not experimental confirmation. Although the nonlinear wall subproblem has been solved with internal convergence checks, independent full-flow benchmarking of the composite approximation and experimental inverse-recovery tests are still required before quantitative use in noncontact rheometry. The resulting carefully delimited free-surface framework also provides a clear starting point for future two-fluid interfacial-wave research.

\appendix
\section{Compact Newtonian eigenfunction check}\label{app:newtonian}
For completeness, the deep-water Newtonian normal mode $e^{i(kx-\omega t)}$ admits the \emph{vertical-velocity-first} representation
\begin{equation}
\begin{aligned}
W(z)&=ce^{kz}+de^{mz},\qquad U(z)=\frac{i}{k}W'(z),\\
\widehat\Pi(z)&=\frac{i\rho\omega c}{k}e^{kz},\\
m^2&=k^2-\frac{i\omega}{\nu},\qquad \Ree(m)>0.
\end{aligned}
\label{eq:Wfirst}
\end{equation}
The clean-surface tangential-stress condition $U'(0)+ikW(0)=0$ gives $(m^2+k^2)d=-2k^2c$, while the kinematic condition gives $\widehat\eta=i(c+d)/\omega$. Substitution into the linear normal-stress condition yields
\begin{equation}
\boxed{\omega^2+2i\nu k^2\omega
\left(1-\frac{2km}{m^2+k^2}\right)
=\Omega^2\frac{m^2-k^2}{m^2+k^2}.}
\label{eq:exactviscous}
\end{equation}
When $\nu k^2/\Omega\ll1$, $m\simeq(1-i)\sqrt{\Omega/(2\nu)}$, and expansion of Eq.~\eqref{eq:exactviscous} returns Eq.~\eqref{eq:newtonian}. The surface tangential strain follows
\begin{equation}
\widehat{2D_{xz}}(z)=\frac{i}{k}[W''(z)+k^2W(z)]
=2ikc(e^{kz}-e^{mz}),
\label{eq:surfacecancel}
\end{equation}
which vanishes exactly at $z=0$. Thus the rotational correction is required to satisfy the Newtonian stress condition even though its correction to the \emph{integrated leading-order} deep-water loss is subdominant in the weak-viscosity limit.

\section{Finite-depth quadrature and wall identities}\label{app:checks}
For $n=1$, $\mean{\sin^2\theta}=\mean{\cos^2\theta}=1/2$ in Eq.~\eqref{eq:J}; hence
\begin{equation}
J_1(H)=\frac1{2\sinh^2H}\int_0^H\cosh(2q)\dd q
=\frac12\coth H.
\label{eq:check1}
\end{equation}
For $H\to\infty$ the surface-centered coordinate $\xi=H-q$ makes the dimensionless strain kernel approach $e^{-\xi}$, giving $J_n\to\int_0^\infty e^{-(n+1)\xi}\dd\xi=1/(n+1)$.

Multiplying Eq.~\eqref{eq:wallPDE} by the velocity \emph{defect} $v$ and integrating by parts gives
\begin{equation}
\frac{\dd}{\dd t}\int_0^\infty\frac{\rho v^2}{2}\dd y
=-v(0,t)\tau_{xy}(0,t)
-\int_0^\infty K|v_y|^{n+1}\dd y.
\label{eq:wallidentity}
\end{equation}
On a periodic solution its averaged storage derivative vanishes; with $v(0,t)=-u_e(t)$, the effective driving term $\mean{u_e\tau_{xy}(0,t)}$ equals the wall dissipation. This term comes from the \emph{defect-flow} equation and is not mechanical work performed by the physically stationary rigid wall. It is a useful independent numerical check when solving for $I_n$.

\section{Dimensional and constitutive consistency}\label{app:notation}
The temporal bulk coefficient has dimension
$[C_t]=\mathrm{m}^{1-n}\mathrm{s}^{-1}$, and the spatial coefficient
$[C_x]=\mathrm{m}^{-n}$. The finite-depth wall coefficient has
$[C_{t,h}^{\rm wall}]=\mathrm{m}^{(1-n)/(n+1)}\mathrm{s}^{-1}$.
Thus every term in Eq.~\eqref{eq:composite} has dimensions of an
amplitude derivative. The quadratures $J_n(H)$ and $I_n$ are
dimensionless; neither is an apparent viscosity. When $n=1$ and
$K=\mu$, the derived bulk equations reduce to
$\dot A=-2\nu k^2A$ and the bottom-layer term reduces to the
standard Newtonian Stokes-layer result.

\bibliographystyle{apsrev4-1}
\bibliography{paper1_references}
\end{document}